\documentclass[useAMS,usenatbib]{mn2e}
\usepackage{color}
\usepackage{aas_macros}
\usepackage{epsfig}
\usepackage{amsmath,amssymb}
\usepackage{multicol}
\usepackage{amsmath,graphicx}
\usepackage{booktabs,caption,fixltx2e}
\usepackage[flushleft]{threeparttable}
\usepackage{multirow}

\usepackage[caption=false]{subfig}

\title[Fluctuations in the EBL]{Effects of Spatial Fluctuations in the Extra Galactic Background Light on Hard Gamma Ray Spectra}
\author[A. M. Kudoda and A. Faltenbacher]{A. M. Kudoda$^{1}$\thanks{E-mail:949916@students.wits.ac.za} and A. Faltenbacher$^{1}$\\
  $^{1}$School of Physics, University of the Witwatersrand, Johannesburg, 2050}
\begin{document}
\date{Accepted   . Received ; in original form }
\pagerange{\pageref{firstpage}--\pageref{lastpage}} \pubyear{2014}
\maketitle
\label{firstpage}
\begin{abstract}
This study investigates the impact of the fluctuations in the extra galactic background light (EBL) on the attenuation of the hard $\gamma$-ray spectra of distant blazars. EBL fluctuations occur on the scales up to 100 Mpc and are caused by clustering of galaxies. The EBL photons interact with high energy $\gamma$-rays via the electron-positron pair production mechanism: $\gamma + \gamma' \rightarrow e^+ + e^-$. The attenuation of $\gamma$-rays depends on their energy and the density of the intervening EBL photon field. Using a simple model for the evolution of the mean EBL photon density, we implement an analytical description of the EBL fluctuations. We find that the amplitudes of the EBL energy density can vary by $\pm 1\%$ as a function of environment. The EBL fluctuations lead to mild alterations of the optical depth or equivalently the transmissivity for $\gamma$-rays from distant blazars. Our model predicts maximum changes  of $\pm 10\%$ in the $\gamma$-ray transmissivity. However, this translates into marginal differences in the power law slopes of currently observed $\gamma$-ray spectra. The slopes of deabsorbed $\gamma$-ray spectra differ by not more than $\pm 1\%$ if EBL fluctuations are included.
\end{abstract}
\section{Introduction}
\label{sec:intro}
The Extragalactic Background Light (EBL) is the light between $0.1$ and $1000$ $\mu$m  accumulated by the global stellar population throughout cosmic time. It contributes about $1\%$ to the sky brightness from the Ultraviolet (UV) to Infrared (IR) wavelength range \citep{Bernstein2000} with two peaks dominating the spectrum: a first peak between $0.1$ and $10$ $\mu$m, which is due to direct stellar emission; and a second peak between $10$ and $1000$ $\mu$m caused by the dust contribution \citep{Hauser2001, Dwek2012}.

The EBL contains a wealth of information related to the evolution and the structure of the Universe and its astrophysical components. Measuring the EBL directly, however, is difficult for various reasons. For example, it is challenging to separate the EBL from the zodiacal light of our solar system and from the foreground light of our Galaxy \citep{Costamante2013}. For direct source counts the foreground light causes systematic bias against the detection of individual faint or low surface brightness galaxies and intergalactic stars. Despite all difficulties, the measurement of the EBL can provide useful integral constraints on star formation models and the baryonic matter content of the Universe \citep{Bernstein2000}.

Several observational studies have been carried  out to investigate fluctuations in the EBL. Fluctuations in the EBL are expected due to the density fluctuations of the underlying galaxy population. For instance, \cite{Shectman1973} and \cite{Shectman1974} study the anisotropy in the optical regime and \cite{Kashlinsky1996part1} investigate the clustering in the near-IR region using Cosmic Background Explorer (COBE) and Diffuse Infrared Background Experiment (DIRBE) Maps. Although one cannot determine the EBL density by only knowing the fluctuations in the EBL, the fluctuation measurements can be used to derive limits on the EBL density at certain regions of the spectrum \citep{Penin2012, Dwek2012}.

Indirect measurements of the EBL can be obtained from observations of the attenuation of $\gamma$-ray spectra of distant very high energy (VHE) sources such as quasars. As $\gamma$-rays travel through the Universe they interact with EBL photons producing electron-positron pairs ($\gamma + \gamma' \rightarrow e^- + e^+$). This process leaves a finger print on the spectral index of the quasars in the VHE regime \citep{Yuan2012}. Electron-positron pair production is the main channel for $\gamma$-ray absorption at VHE, and the only absorption process we consider here for $\gamma$-ray photons traveling cosmic distances. It can take place only if the total energy of the two interacting photons is higher than the rest mass of the electron-positron pair. Based on the cross section of the photon-photon interaction and a model of the EBL density the optical depth ($\tau$) can be computed for VHE $\gamma$-ray sources at redshift $z$.

It is well known that galaxies, the sources of the EBL, are clustered. Thus, the EBL density is expected to be subject to fluctuations too. However, since the EBL is accumulated throughout the history of the Universe the fluctuations may be small as they are largely overpowered by the homogeneous contribution of EBL sources at larger distances. To what extent these small fluctuations can impact the attenuation of VHE quasar spectra is currently still under debate. Recently, \cite{Furniss2015} found a correlation between VHE-emitting sources and cosmic voids along the line of sight. However, they estimate that the attenuation decreases less than $10\%$ for a $\gamma$-ray source with opacity $\tau \sim 5$ if the line of sight goes entirely through under-dense regions. The model presented here provides a quantitative analysis of EBL fluctuations which allows us to predict the alteration of the attenuation of VHE $\gamma$-rays from distant quasars.

The structure of this paper is set as follows. In Section~\ref{sec:mod} we introduce the analytical EBL model which is based on the work of \cite{Razzaque2009} and \cite{Finke2010} with an extension allowing for the implementation of EBL fluctuations. The calculations for the absorption of $\gamma$-rays due the interaction with EBL photons are reviewed in Section~\ref{sec:abs}. Based on deabsorbed quasar spectra the impact of the EBL fluctuations is discussed in Section~\ref{sec:res}. Finally, a brief conclusion is given in Section~\ref{sec:con}.
\begin{figure}
  \begin{center}
    \includegraphics[width=0.5\textwidth]{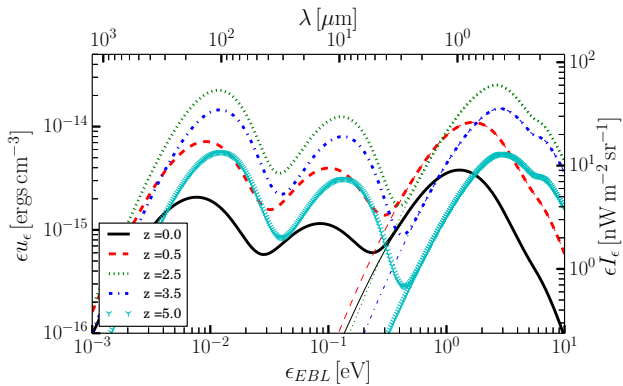}
    \caption{Model of the EBL energy density including stellar and dust components at different redshifts (solid lines). The model of the stellar component (dashed lines) is a reproduction of model B in \protect\cite{Razzaque2009}. The emission of the dust component, responsible for the two peaks at $\lambda \gtrsim 8{\rm \mu m}$, is computed following \protect\cite{Finke2010}.}
    \label{fig:R}
  \end{center}
\end{figure}
\section{Modelling the EBL}
\label{sec:mod}
The EBL model employed here is similar to the approach presented in \cite{Razzaque2009}, hereafter referred to as RDF09, which is a forward evolution model. RDF09 only takes into account the contribution of the stellar component which is inferred from the stellar thermal surface emission of main-sequence stars modelled as blackbody radiation. Contributions from off-main sequence stars are ignored. Since the total contribution of quasars and active galactic nuclei (AGNs) to the optical and the infrared regime of the EBL spectrum is $10\%-20\%$ at maximum \citep{Hauser2001} the contributions of AGNs are ignored by RDF09 and in our model as well.

The IR re-emission of stellar light absorbed by dust is modeled following \cite{Finke2010}, hereafter referred to as FRD10. As an extension we implement a statistical description of fluctuations of the EBL photon densities due to the clustering of galaxies. For that purpose we utilise the small scale ($<$ 100 Mpc) variance of the cosmic star formation rate.
\subsection{Stellar Component}
\label{ssc:ste}
This section gives a brief review of the RDF09 model, for more details we refer the interested reader to the original article by \cite{Razzaque2009}. The RDF09 model estimates the EBL at $z = 0$ by integrating the contributions from stars of all masses formed throughout the history of the universe. The emission of stars is modeled as blackbody radiation at a given temperature emitted from the spherical surface of the star. Temperature and radius are expressed as fitting functions with the stellar mass as free variable. The mass distribution is modeled by a given initial mass function (IMF) plus a stellar life time criterion which eliminates the contribution of stars that have left the main sequence. A further element for the integration of the EBL is the global star formation rate (SFR). 

With these ingredients the spectral energy distribution (SED) of the stellar component of the EBL measured at present ($z = 0$) can expressed by: 
\begin{align}\label{eq:R}
  \frac{dN(\epsilon,z=0)}{d\epsilon dV}= &\mathcal{N} \int^\infty_{z=0} dz'' \left|\frac{dt}{dz''}\right| \psi(z'') \int^{M_{\text{max}}}_{M_{\text{min}}} dM \xi(M) \nonumber\\ 
  &\times\int^{z''}_{\text{max}\{z=0,z_d(M,z')\}}dz' \left|\frac{dt}{dz'}\right|\nonumber\\ 
  &\,\,\, f_{\rm esc}(\epsilon') \frac{dN(\epsilon',M)}{d\epsilon'dt}(1+z'), 
\end{align}
where $\mathcal{N}$ is the normalization factor for the IMF $\xi(M)$, $M_{min}$ and $M_{max}$ are $0.1 \rm ~ M_\odot$  and $120 \rm ~ M_\odot$, $\psi(z)$ is the SFR at redshift $z$, $\epsilon'=\epsilon(1+z')$ is the redshifted energy of the EBL photons, $z_d$ is the redshift when the star evolves away from the main-sequence, and $\frac{dN(\epsilon',M)}{d\epsilon'dt}$ is the total number of emitted photons per time per energy intervals from a star with radius $R$ and temperature $T$. The averaged photon escape fraction from a galaxy, $f_{\rm esc}(\epsilon')$, is given by an empirical fitting function adapted from \cite{Driver2008}. The fraction of photons generated in stars which does not escape is $1-f_{\rm esc}(\epsilon')$. In the RDF09 model these photons are considered to be lost. In the FRD10 model, discussed next section, these photons are re-emitted in the IR band. 

RDF09 considers different combinations between the SFR and IMF. We use their best fitting EBL model (model-B with single power law for the mass-luminosity and mass-temperature relations) to calculate the SED. Model-B is based on the SFR described in \cite{Cole2001}\footnote{$\psi(z) = [h(a+bz)]/[1+(z/c)^d]$ with fitting parameters from \cite{Hopkins2006}: \{$a$, $b$, $c$, $d$\}=\{0.0166, 0.1848, 1.9474, 2.6316\} } and the modified  ``Salpeter A" IMF, $dN/dM \propto M^{-\kappa}$ with $\kappa = 1.5$ for $M<0.5M_\odot$ and $\kappa=2.35$ for $M>0.5M_\odot$. 

In the following we will make use of the comoving EBL energy density, $\epsilon u_{\epsilon}$, at a given redshift $z = z_1$ which can be derived from the SED (Eq.~\ref{eq:R}) as follows:
\begin{equation}
  \epsilon u_{\epsilon} = (1 + z_1)^4 \epsilon^2 \frac{dN(\epsilon,z=z_1)}{d\epsilon dV}\ .
\end{equation}
Note, for the determination of the comoving EBL energy density at $z = z_1$ all occurrences of ``$z=0$'' in Equation~\ref{eq:R} have to be replaces by ``$z=z_1$''. Figure~\ref{fig:R} shows the evolution of $\epsilon u_{\epsilon}$ with redshift. The dashed lines represent the stellar component and the solid lines show the full EBL with dust re-emission which is discussed in the following section.  
\begin{figure}
  \begin{center}
    \includegraphics[width=0.5\textwidth]{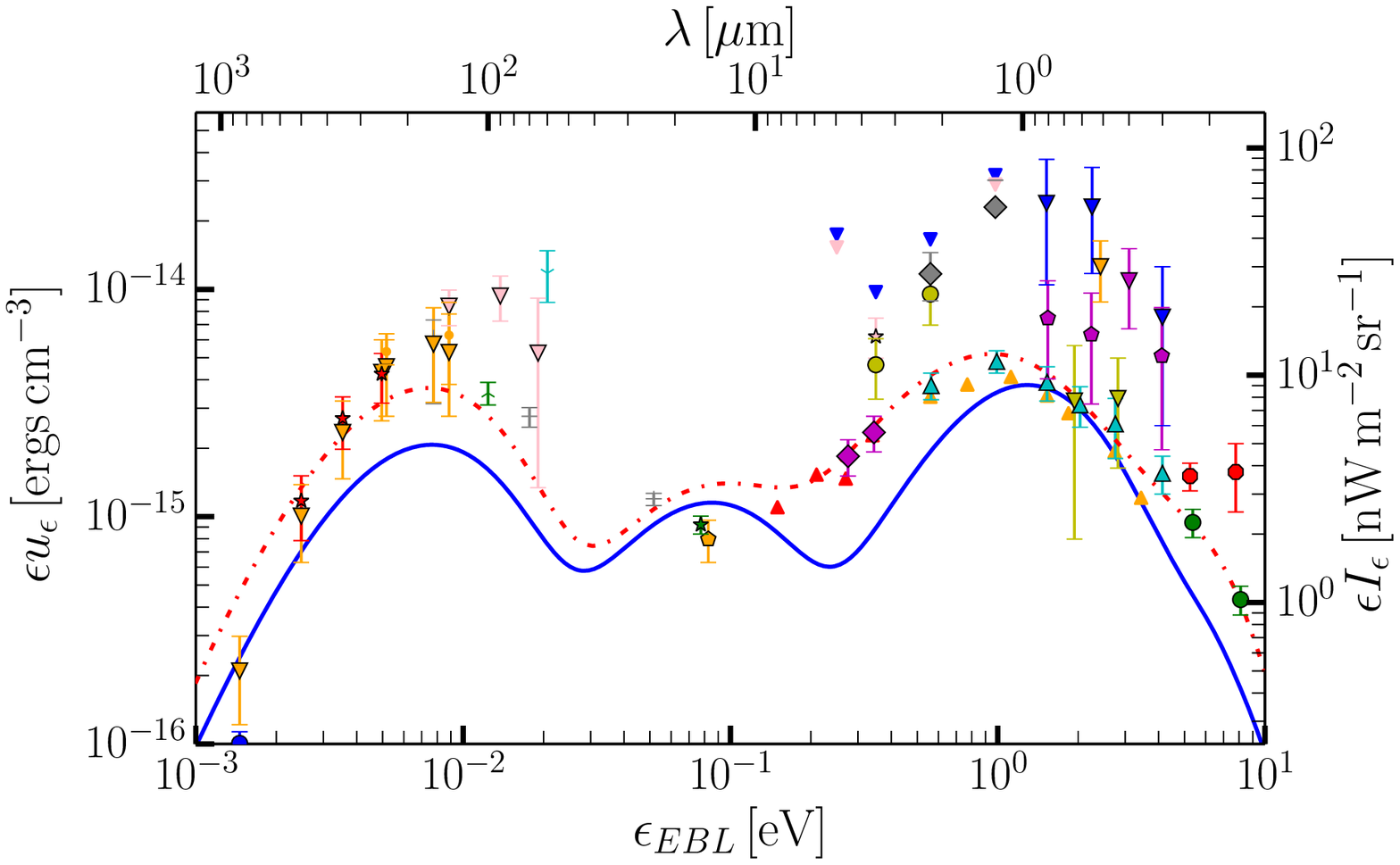}
    \caption{Comparison of EBL energy density model spectra with observational data. The solid black line represent our EBL energy density at $z=0$ and the dashed red line is model by \protect\cite{Finke2010}. The symbols represent EBL measurements: 
green circle - \protect\cite{Xu2005}; 
red octagon   - \protect\cite{Gardner2000}; 
blue triangle down  - \protect\cite{Bernstein2007}; 
cyan triangle up   - \protect\cite{Totani2001};  
magenta triangle down  - \protect\cite{Mattila1990}; 
yellow triangle down  - \protect\cite{Matsuoka2011}; 
orange triangle down  - \protect\cite{Dube1979}; 
grey triangle down  - \protect\cite{Levenson2007}; 
pink triangle up  - \protect\cite{Keenan2010}; 
magenta diamond  - \protect\cite{Ashby2013}; 
blue square   - \protect\cite{Arendt2003}; 
orange pentagon  - \protect\cite{Hopwood2010}; 
green star  - \protect\cite{Teplitz2011}; 
grey plus  - \protect\cite{Bethermin2010}; 
cyan triangle down  - \protect\cite{Finkbeiner2000}; 
pink triangle down  - \protect\cite{Matsuura2011}; 
green triangle up  - \protect\cite{Berta2011}; 
orange triangle down  - \protect\cite{Fixsen1998}; 
grey triangle down  - \protect\cite{Penin2012}; 
red star  - \protect\cite{Bethermin2012}; 
orange dot  - \protect\cite{Odegard2007}; 
blue circle  - \protect\cite{Zemcov2010}.}
    \label{fig:EBLWithData}
  \end{center}
\end{figure}
\begin{table}
  \caption{Dust Parameters}
  \label{tab:dust}
  \begin{center}
    \begin{tabular}{ |c|c|c|c| }
      \hline\hline
      Component & $n$ & $f_n$ &  $T_n(\rm K)$ \\
      \hline
      Warm large grains & 1 & 0.60 &  40\\
      Hot small grains  & 2 & 0.05 &  70\\
      PAHs              & 3 & 0.35 & 450\\
      \hline
    \end{tabular}
  \end{center}
\end{table}
\subsection{Dust Component}
\label{ssc:dus}
In order to compute the dust re-emission spectrum, we start with the total number of photons emitted per unit energy and time by a star of radius $R$ is given as discussed in RDF09: 
\begin{equation}
  {dN \over {d\epsilon  dt}} = {{R^2} \over {\pi c^2 \hbar^3 }} {{\epsilon^2}\over{\exp(\epsilon/kT) - 1}}\ . 
  \label{eq:star}
\end{equation} 
RDF09 present fitting formulae which allow to compute all relevant stellar quantities, such as radius $R$, temperature $T$ and luminosity $L$ as functions of stellar mass $M$. The averaged fraction of photons absorbed by the dust is given by $1 - f_{\rm esc}$ (cf. Eq.~\ref{eq:R}). The absorbed photons are re-radiated in the IR. Assuming quasi static equilibrium, the total energy absorbed and re-emitted by the dust per unit time interval, $dE_{\rm dust}(M) / dt$, for a star of mass $M$ is given by:
\begin{equation}
  {dE_{\rm dust}(M) \over {dt}} = \int_0^\infty \ (1 - f_{\rm esc}) \ \epsilon \ {{dN(\epsilon,M)}\over{d\epsilon dt}} \ d\epsilon\ . 
  \label{eq:dust}
\end{equation} 

To compute the dust IR emission spectrum the interstellar medium is modeled by three major dust components \citep[cf.][]{Desert1990}: (1) large dust grains; (2) small dust gains; and (3) polycyclic aromatic hydrocarbons (PAHs). The superposition of the three associated blackbodies results in the dust IR emission spectrum. The properties of the three blackbodies are indicated in Table~\ref{tab:dust}, where $n$ is an arbitrary index, $f_{\rm n}$ is the fraction of absorbed emissivity re-radiated by a particular blackbody at temperature $T_{\rm n}$. Consequently, the total absorbed emissivity is equal to: 
\begin{equation}
  {dE_{\rm dust}(M) \over {dt}} = \sum_{n=1}^3 \int_0^\infty \ {{\Omega_n(M) \epsilon^3}\over{\exp(\epsilon/kT_n)}} d\epsilon\ , 
  \label{eq:dust}
\end{equation} 
where the scaling factors $\Omega_n(M)$ can be determined as follows:
\begin{equation}
  \Omega_n(M) = f_n
  {{dE_{\rm dust}(M) /dt} \over {\int_0^\infty  {{\epsilon^3}\over{\exp(\epsilon/kT_n)}} d\epsilon}}\ .
  \label{eq:omega}
\end{equation} 

The dust re-emission term (Eq.~\ref{eq:dust}) is plugged into Equation~\ref{eq:R} resulting in the following integral expression for the comoving EBL energy density measured at redshift $z = z_1$: 
\begin{align}\label{eq:Rplus}
  \frac{dN(\epsilon,z=z_1)}{d\epsilon dV}= &\mathcal{N} \int^\infty_{z=z_1} dz'' \left|\frac{dt}{dz''}\right| \psi(z'') \int^{M_{\text{max}}}_{M_{\text{min}}} dM \xi(M) \nonumber\\ 
  &\times\int^{z''}_{\text{max}\{z=z_1,z_d(M,z')\}}dz' \left|\frac{dt}{dz'}\right| (1+z') \nonumber\\ 
  & \left[ \sum_{n=1}^3 {{\Omega_n(M) \epsilon'^2}\over{\exp(\epsilon'/kT_n)}} + f_{esc}(\epsilon') \frac{dN(\epsilon',M)}{d\epsilon'dt} \right] 
\end{align}
The solid lines in Figure~\ref{fig:R} show the comoving EBL energy densities for a set of redshifts as indicated. Figure~\ref{fig:EBLWithData} compares the EBL spectrum for the present epoch (blue solid line) with observations (symbols) and the FRD10 model (red dashed line). The FRD10 model comprises a more sophisticated stellar evolution model (based on fits to stellar evolution tracks presented in \citealt{Eggleton1989}), which leads to a slightly higher integrated luminosity of individual stars. Consequently, our model provides a lower limit to the EBL measurements as it employs a slightly underluminous stellar population. This has no effect on our results as we are interested in the \emph{relative differences} of the $\gamma$-ray attenuation caused by the fluctuations of the EBL photon densities.

\begin{figure}
   \begin{center}
      \includegraphics[width=0.5\textwidth]{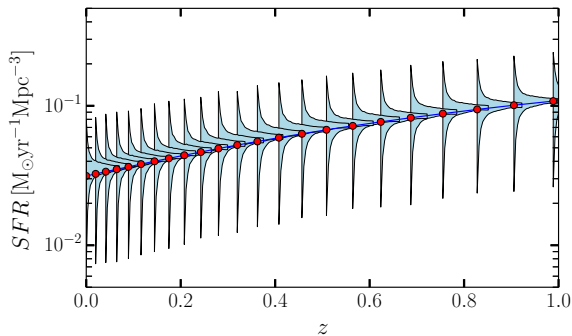}
      \caption{The SFR history derived from the semi-analytical galaxy catalogue by \protect\cite{Guo2013} as function of redshift. The red points represent the mean star formation rate in the entire simulation box at the redshifts of the snapshots. The fluctuations are indicated as vertical, ``trumpet-shaped'', shaded regions delineated by the lower $5$ and upper $95$ percentiles of the distribution of SFRs in shells of various radii (redshifts) about 1000 centres located on a regular grid within the simulation box.}
      \label{fig:sfr}
   \end{center}
\end{figure}
\subsection{Implementing Spatial EBL Fluctuations} 
\label{ssc:new}
Galaxies show spatial clustering which causes spatial fluctuations in the intensity of the EBL. In order to develop a mathematical framework for computing the EBL fluctuations we start by determining the spatial fluctuations in the SFR. The rationale behind this approach is that: (1) the calculation of the EBL in the RDF09 model is based on the integration of the SFR over cosmic time and (2) spatial differences in the SFR cause spatial differences in the galaxy density. Thus, if a region shows a lower-than-average SFR then the galaxy density is lower and thus also the EBL intensity is lower compared to the cosmic mean. For regions of higher-than-average SFR the opposite holds true.

For a rough (but sufficient) estimate of the spatial fluctuations in the SFR we employ a semi-analytical galaxy catalogue based on the Millennium Simulation Run $7$ (MR7) \citep{Lemson2006,Guo2013}, which reproduces the observed galaxy populations reasonably well. We extract the distribution of the SFR within shells of 10 Mpc/h thickness about 1000 seed points on a regular grid filling the volume of the simulation box. The 5th and 95th percentiles of the distribution of SFRs in each shell give a measure of the scatter in the SFR as a function of distance to the location of the observer.

The determination of the scatter is repeated for all snapshots which comprise galaxies ($z \lesssim 10$). The comoving shell radii, $R_{\rm i}$, can be converted into a redshift separation, $z_{\rm i} = z - z_{\rm snap}$, by using the distance-redshift relation, $c z_{\rm i} = H(z_{\rm snap})\ a(z_{\rm snap})\ R_{\rm i}$, where $z_{\rm snap}$ is the redshift of the snapshot, $R_{\rm i}$ is the comoving distance, $H(z)$ and $a(z)$ are the Hubble parameter and the scale factor at that redshift. Figure~\ref{fig:sfr} shows the mean SFR in the simulation box for the snapshots between $z=0$ and $z=1$ (red circles) and the scatter as a function of $z$  (blue `trumpets'). The trumpet shape is easily explained, locally within $R_{\rm i}\lesssim 100$ Mpc/h ($z_{\rm i} \lesssim 0.03$) the scatter in the SFR is large, but as the radii of the shells increase the scatter of the SFR reduces, which is simply a volume effect. To obtain a time-independent approximation for the spatial scatter of the SFR we fit polynomials to the trumpet shapes of all redshifts simultaneously: one polynomial for the upper 95 percentile margin and another polynomial for the lower 5 percentile margin.

Adding the upper or lower margins to the smooth SFR at a given redshift allows us to model the EBL intensity in over- and under-dense regions, respectively. The continuous use of the upper margin of the SFR for the computation of the local EBL intensity along the path of a $\gamma$-ray models a $\gamma$-ray traversing the Universe within a cylinder of enhanced galaxy (SFR) densities. On the contrary, a continuous use of the lower margin of the SFR mimics a $\gamma$-ray within a cylinder of reduced galaxy (SFR) densities. These are the two extreme cases which we employ to determine the upper and lower limits of $\gamma$-ray attenuation.
\begin{figure}
  \begin{center}
    \includegraphics[width=0.5\textwidth]{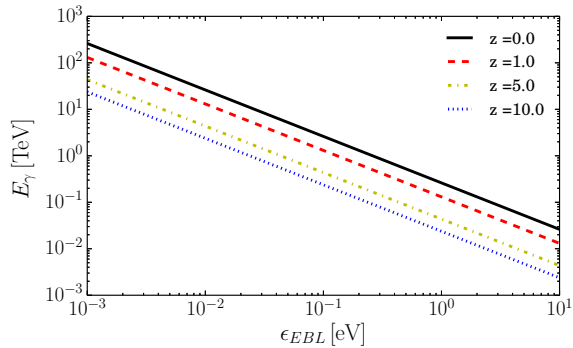}
    \caption{Limits for $\gamma + \gamma' \rightarrow e^- e^+$ pair production for various redshifts. The lines represent the $s_0 = 1$ condition \protect\citep{Gould1967} as a function of $\gamma$-ray, $E_\gamma$ and EBL photon energy, $\epsilon_{EBL}$. For each given redshift pair production is possible in the area above the line.}
    \label{fig:Gam}
  \end{center}
\end{figure}
\begin{figure}
  \begin{center}
    \includegraphics[width=0.5\textwidth]{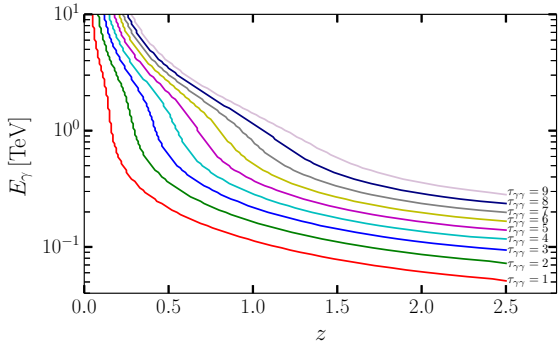}
    \caption{$E-z$ contours for values of $\tau_{\gamma\gamma} = 1 - 9 $ based on the EBL energy density predicted by our model. Each line is labeled by the corresponding $\tau$. The bottom line for $\tau_{\gamma\gamma} = 1$ defines the $\gamma$-ray horizon which separates $\gamma$-ray transparent (below) and $\gamma$-ray opaque (above) areas.}
  \label{fig:hor}
  \end{center}
\end{figure}
\begin{figure}
  \begin{center}
    \includegraphics[width=0.5\textwidth]{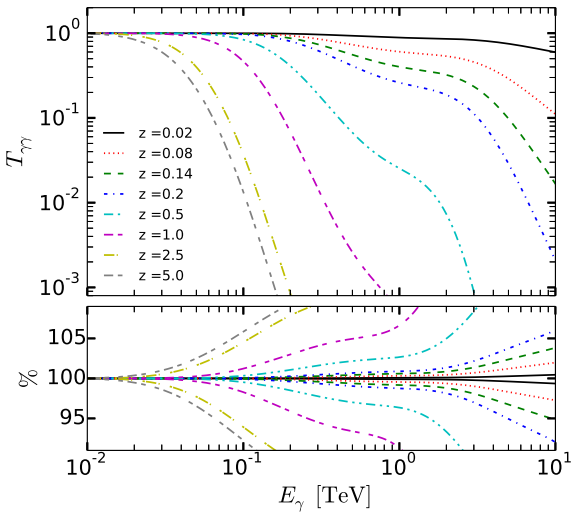}
    \caption{\emph{Upper panel}: $gamma$-ray transmissivity as a function of $\gamma$-ray energy at different redshifts. \emph{Lower panel}: The relative difference between the mean $\gamma$-ray transmissivity and the upper and lower limits due to EBL fluctuations as a function of observed $\gamma$-ray energy. The lines above and below the horizontal $100\%$ demarcation correspond to the lower and upper limits of the EBL fluctuations, respectively.}
    \label{fig:tra}
  \end{center}
\end{figure}
\section{Absorption of $\gamma$-ray Spectra}
\label{sec:abs}
The total cross-section, $\sigma$, for electron pair production, $\gamma+\gamma \rightarrow e^+ + e^-$ \citep[cf.][]{Gould1967} is:
\begin{align}\label{eq:cro}
 \sigma = \frac{1}{2} \pi r_0^2 (1-\beta^4) \left[ (3+\beta^4) \ln \frac{1+\beta}{1-\beta}-2\beta (2-\beta^2) \right]\, 
\end{align}
where $\beta = \sqrt{1-\frac{1}{s}}$ is the electron and positron velocity, $r_0$ is electron radius, and  $s = \frac{s_0}{2} (1-\cos\theta)$ is square energy of the centre of mass, where $s_0 = \left(\frac{\epsilon E}{m^2c^4}\right)$. According to Eq.~\ref{eq:cro} pair production can only occur if $s>1$. Figure~\ref{fig:Gam} shows the $s=1$ lines as function of the $\gamma$-ray and EBL photon energies at different redshifts. Pair production is possible in the regions above the lines. At $z=0$, for instance, $\gamma$-rays above $\sim 261$ TeV can produce $e^-,e^+$ pairs through interactions with EBL photons of the entire range shown ($10^{-3}$ eV $ \lesssim \epsilon_{EBL} \lesssim 10$ eV). On the contrary, the EBL is transparent to the $\gamma$-ray photons below $\sim 0.02$ TeV. 

Based on the total cross section, Eq~\ref{eq:cro}, we calculate the optical depth, $\tau_{\gamma\gamma}$, as a function of $\gamma$-ray energy, $E$, and redshift, $z$ \citep[cf.][]{Razzaque2009}:   
\begin{align}\label{eq:opt}
   \tau_{\gamma\gamma}(E_\gamma,z) = c \pi r_0^2\left(\frac{m_e^2c^4}{E_\gamma}\right)^2 &\int^z_0 \frac{dz_1}{(1+z_1)^2}\left| \frac{dt}{dz_1}\right| \nonumber\\ 
    \times &\int^\infty_{m_e^2c^4/E_\gamma(1+z_1)} d\epsilon_1\frac{u_{\epsilon_1}}{\epsilon_1^3} \overline{\varphi}[s_0(\epsilon_1)]\ ,
\end{align}
where $s = E_\gamma (1+z_1) \epsilon_1 / 2m_e^2c^4$, $\varphi[s_0(\epsilon_1)] = \int^{s_0(\epsilon)}_1 s \overline{\sigma}(s)ds$,  and $\overline{\sigma}(s) = \frac{2\sigma(s)}{\pi r_0^2}$, $s_0 = E_\gamma(1+z_1) / m_e^2c^4$.

As expected $\tau_{\gamma\gamma}(E_\gamma,z)$ increases with $z$, i.e. the $\gamma$-ray emission from a source at larger distance will be attenuated more strongly. For nearby sources $\tau$ is usually $< 1$ in which case the Universe becomes optically thin for $\gamma$-rays. Still, at very high $\gamma$-ray energies the optical depth can be larger than 1,  leading to an optically thick Universe. For a given $E_\gamma$ the redshift at which  $\tau_{\gamma\gamma} = 1$ defines the $\gamma$-ray horizon \citep{Fazio1970}. Figure \ref{fig:hor} shows the $E-z$ contours for values of $\tau_{\gamma\gamma} = 1 - 9 $ based on the EBL energy density predicted by our model. 

The transmissivity, $T_{\gamma \gamma}$ is defined as the probability to observe $\gamma$-rays with energy, $E_\gamma$, from a source at redshift $z$. It can be calculated by:
\begin{align}\label{eq:tra}
 T_{\gamma \gamma}(E_\gamma,z)=e^{-\tau_{\gamma\gamma}(E_\gamma,z)}\ .
\end{align}
The upper panel in Figure \ref{fig:tra} shows the transmissivity as predicted by our EBL model for different redshifts. The graphs in lower panel are discussed in the results section. 
\section{Results}
\label{sec:res}
We use the EBL spectrum discussed in section \ref{sec:mod} as our fiducial model for the mean EBL density. Upper and lower EBL density limits are implemented as discussed in section \ref{ssc:new} corresponding to $\gamma$-rays traveling through over- and under-dense regions along the entire path from the source to the observer. We first discuss the impact of the spatial fluctuations of the SFR on the EBL spectrum. The results are used to determine the mean, the upper and the lower limits of the $\gamma$-ray transmissivity which in turn is employed to deabsorb VHE spectra of observed blazars. At the end of this section we present a speculation on how the spectra of very high redshift blazars would be affected by the fluctuations in the EBL.   
\subsection{Fluctuations in the SFR and EBL}
\label{ssc:flu}
Equation ~\ref{eq:Rplus} is the core of the model presented here. Basically, it integrates over the cosmic SFR to obtain the comoving EBL energy density at $z=z_1$. The SFR is subject to spatial fluctuations which we include in a statistical manner as illustrated in Figure~\ref{fig:sfr}. The integration of the upper and lower limits of the SFR  allows us to compute the EBL intensity in over- and under-dense regions respectively. We find that the EBL intensity changes by $\pm 1\%$ over the entire frequency range depending on whether the observer is located in an over-dense or under-dense region. This relatively small change is expected since the majority of EBL photons arrive from shells with large radii which reflect the average luminosity of the universe. However, cumulative effects for $\gamma$-rays traversing the Universe along predominantly over- or under-dense paths may still alter the $\gamma$-ray attenuation. The following section investigates whether cumulative effects are significant.  
\subsection{Fluctuations in the $\gamma$-rays Transmissivity}
\label{ssc:tra}
The lower panel of Figure~\ref{fig:tra} displays the relative difference between the upper and the lower limits of the transmissivities obtained by plugging in the lower and upper limits of the EBL density in Eq.~\ref{eq:opt}. Choosing the upper (lower) SFR limit for every integration step mimics the cumulative effect of $\gamma$-rays traveling through over  (under) - dense regions along the entire path from the source to the observer. It turns out that spatial fluctuations of the EBL intensity have negligible impact on the transmissivity for $\gamma$-ray sources with $z < 0.5$ and $E_\gamma < 1$ TeV ($\lesssim 2\%$). But for larger energies and redshifts we find increasing deviations from the mean transmissivity. For instance, sources at $z > 0.5$ with $\gamma$-ray energies $E_\gamma> 0.1$ TeV show a change of up to $\pm 10\%$ in the transmissivity. 
\begin{figure*}
  \mbox{ 
    \includegraphics[width=0.5\linewidth]{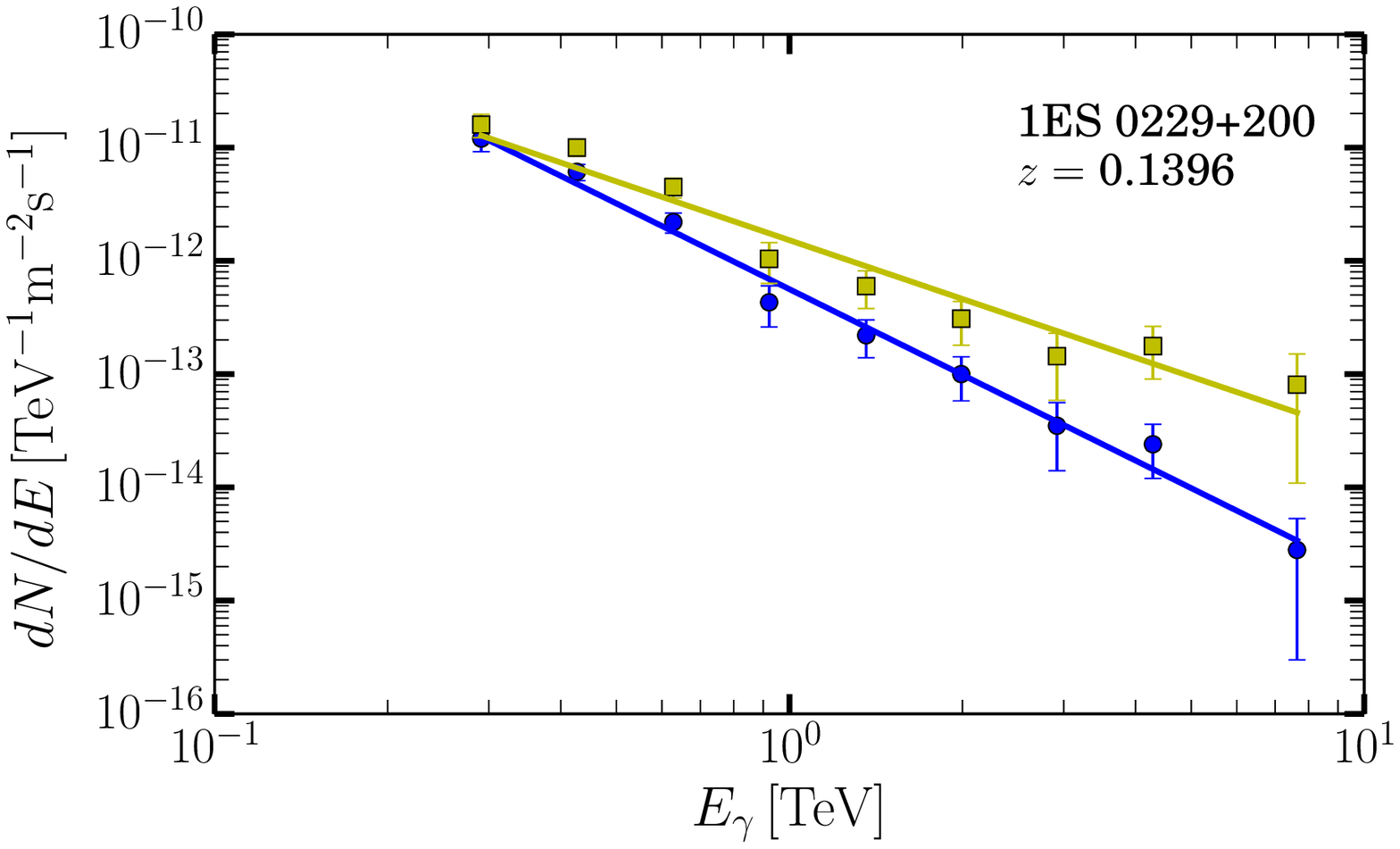}
    \includegraphics[width=0.5\linewidth]{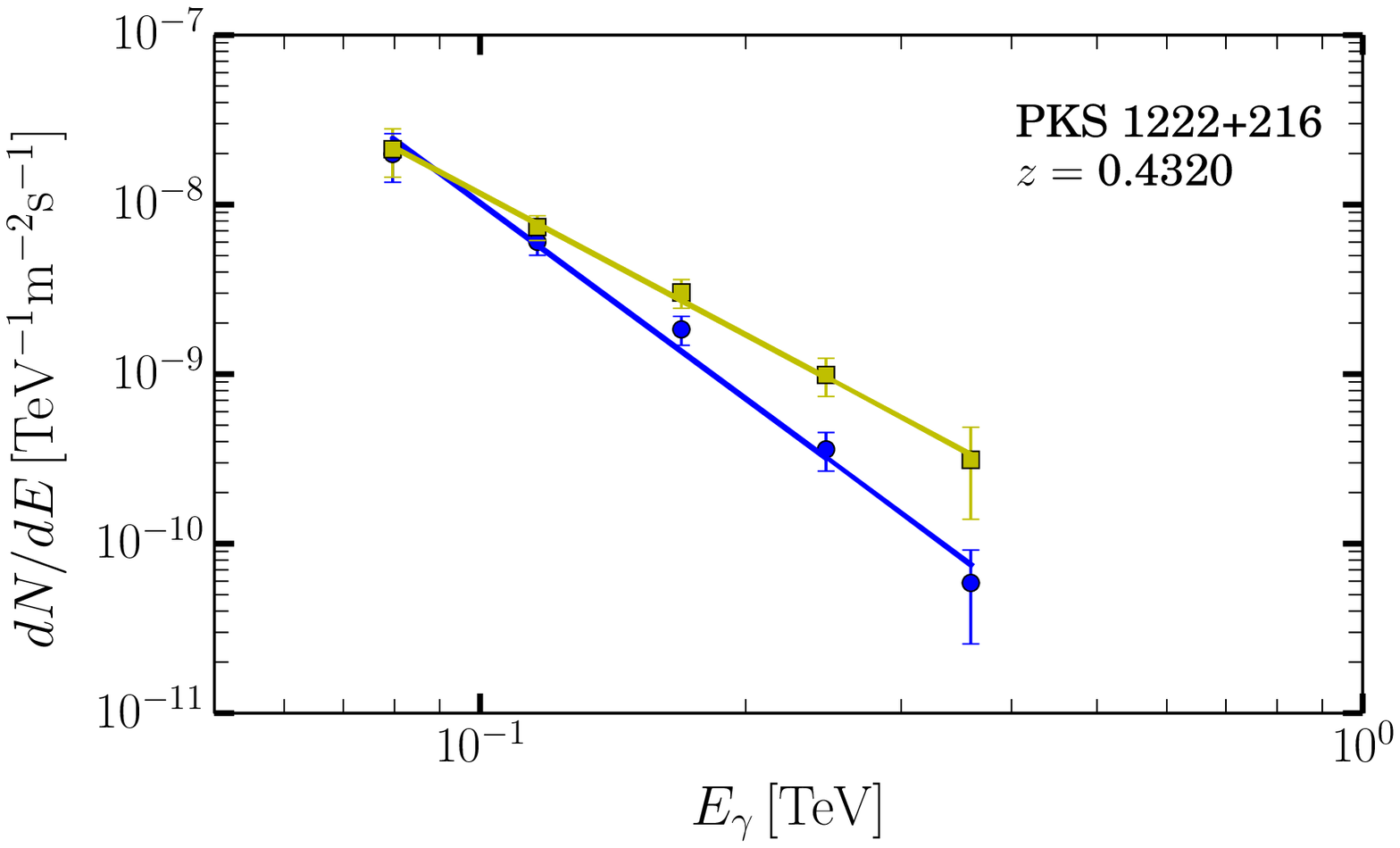}
  }
  \mbox{
    \includegraphics[width=0.5\linewidth]{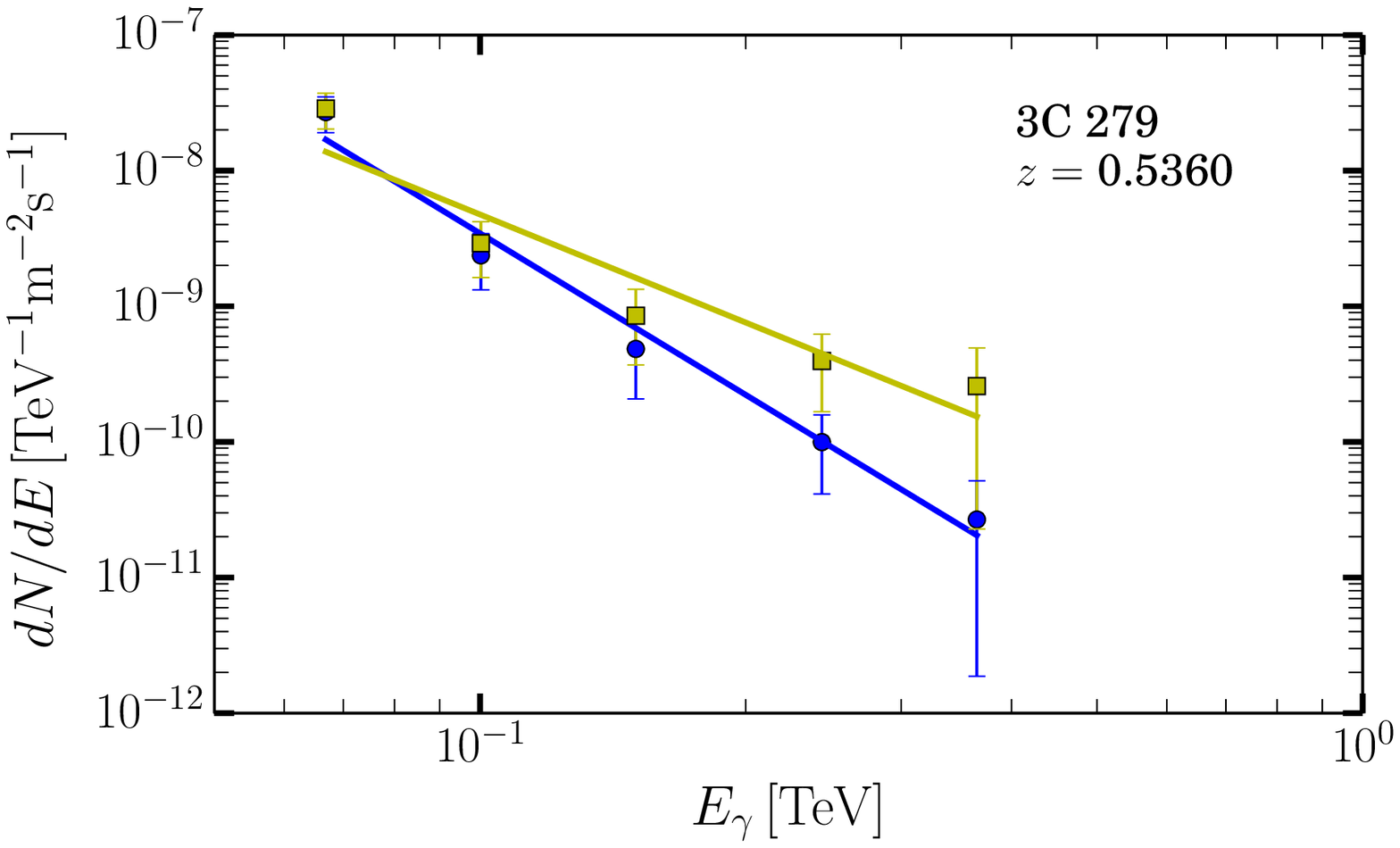}
    \includegraphics[width=0.5\linewidth]{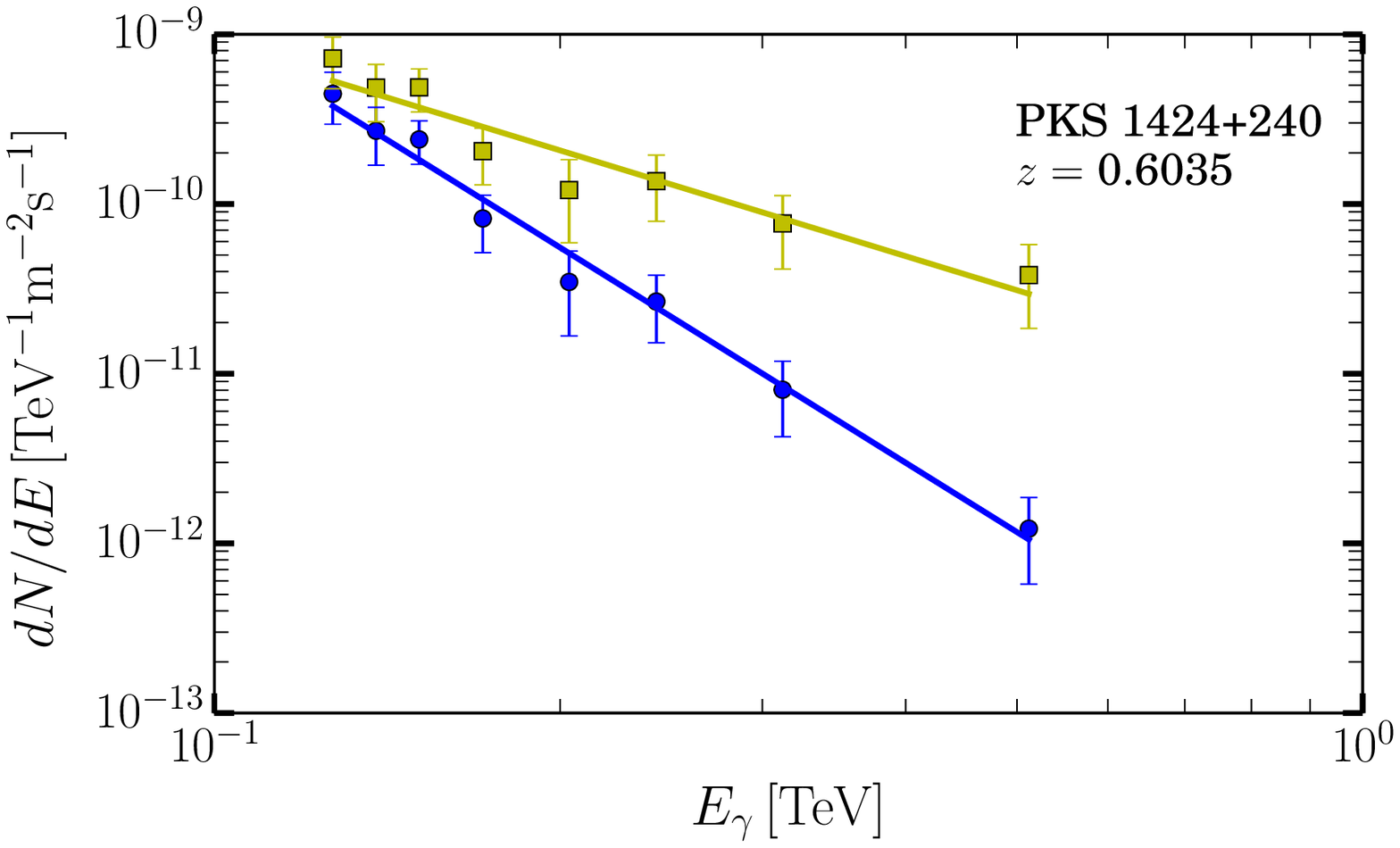}
  }
  \caption{Observed $\gamma$-ray fluxes of blazars indicated as blue circles and deabsorbed counterparts shown as yellow squares. The blue and yellow lines are power law fits ($dN/dE\propto E^{-\Gamma}$) to the observed and the deabsorbed data points, respectively. The $\Gamma$s as obtained from the power law fits are given in Table~\ref{tab:slop}.}
  \label{fig:slop}
\end{figure*}
\begin{table}
  \caption{Logarithmic slope of deabsorbed blazar spectra}
  \label{tab:slop}
  \begin{center} 
    \begin{tabular}{|c|c|c|c||}
      \hline
      Blazar      			& Redshift	& Observed $\Gamma$ 	& Deabsorbed $\Gamma$                      \\
      \hline
      PKS $1424+240^\ddagger$	&  0.6035	&  4.26 		&  $2.07^{2.05}_{2.08}$ \vspace{0.1cm}     \\
      PKS $1222+216^\dagger$ 	&  0.432 	&  3.84 		&  $2.77^{2.75}_{2.78}$ \vspace{0.1cm}     \\
      lES $0229+200^\star$    	&  0.1396	&  2.51 		&  $1.72^{1.71}_{1.73}$ \vspace{0.1cm}     \\
      3C $279^\bullet$       	&  0.536 	&  3.94 		&  $2.66^{2.64}_{2.67}$ \vspace{0.1cm}     \\
      \hline
    \end{tabular}   
  \end{center}  
  \begin{small}     
  \raggedright{$^\ddagger$\protect\cite{Archambault2014}, $^\dagger$\protect\cite{Kushwaha2014}, $^\star$\protect\cite{Aharonian2007} and $^\bullet$\protect\cite{Albert2008} }
  
  \emph{Note}: the super- and sub-scripts in column $4$ represent upper and lower $\Gamma$s based on the upper and lower transmissivity limits of the EBL fluctuation model.
  \end{small}
\end{table}
\subsection{Impact on Deabsorption of $\gamma$-ray Spectra}
\label{ssc:obs}
Knowing the transmissivity allows to determine the deabsorbed $\gamma$-ray flux of the observed VHE blazar spectra: 
\begin{equation}
  \left[ dN \over dE \right]_{\rm deabsorbed} = \exp[\tau(z,E)] \left[ dN \over dE \right]_{\rm observed}
\end{equation}
Deabsorbed spectra provide valuable insights into the intrinsic blazar physics. Figure~\ref{fig:slop} shows the deabsorbed spectra for a sample of four different VHE sources at different redshifts. The lines present power law fits to the observed (blue) and deabsorbed (green) data points. Table~\ref{tab:slop} gives the exponents of the deabsorbed spectra which are slightly larger (spectra are steeper) than reported in the literature. This is a result of the slightly underluminous stellar population used in our model.

Employing the upper and lower limits of the transmissivity due to fluctuations in the EBL causes marginal differences in the deabsorbed data points. The ranges of the exponents of the respective power law fits are indicated in the 4th column of Table~\ref{tab:slop}. The exponents differ by $\sim 1\%$. This leads to the conclusion that fluctuations in the EBL intensity have marginal impact on the slope of the spectra for the redshift range covered by the blazar sample considered here.  
\begin{figure}
  \mbox{\includegraphics[width=0.5\textwidth]{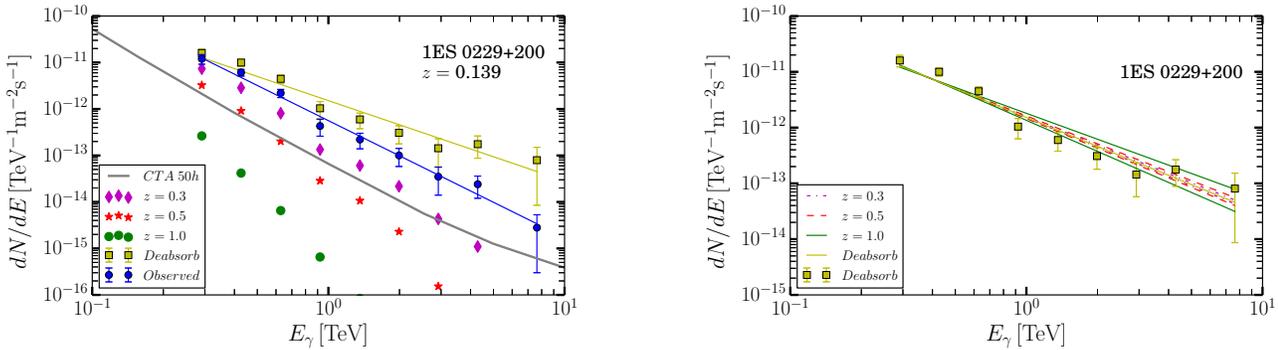}}
  \caption{The blue circles and yellow squares represent observed and the deabsorbed $\gamma$-ray fluxes for $PKS 1222+216$. The lines are power law fits to the data points. The magenta, green and red dots would be the observed fluxes if the source were located at $z=0.3$, $0.5$, and $1.0$, respectively. The grey solid line is the predicted CTA sensitivity for 50 hour observation. It is interesting to note that the slope of the quasars and the sensitivity limit of CTA are very similar.}
  \label{fig:CTA}
\end{figure}
\begin{figure}
  \mbox{\includegraphics[width=0.5\textwidth]{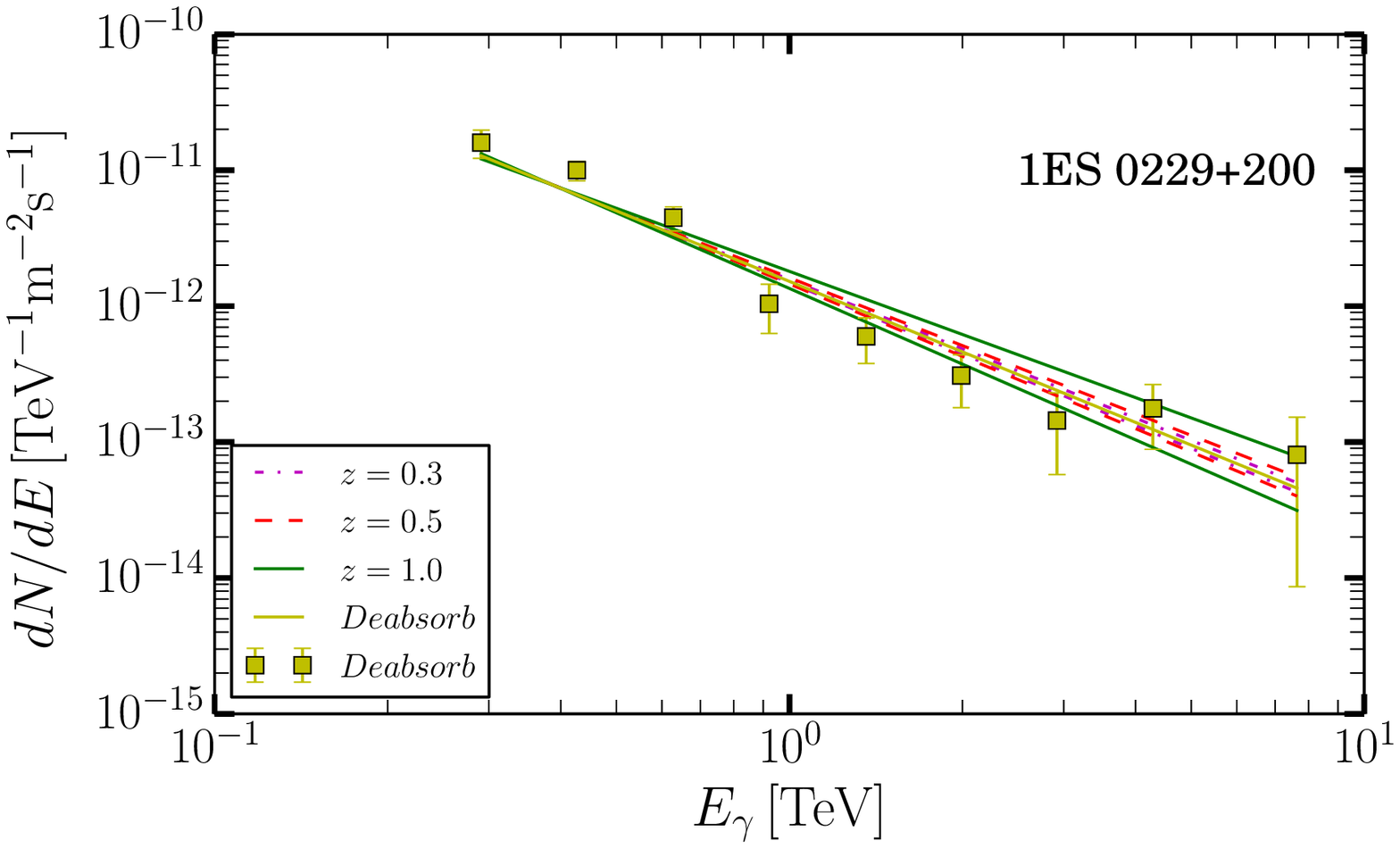}}
  \caption{The yellow squares show the deabsorbed $\gamma$-ray fluxes with measurement errors for PKS $1222+216$. The blue, green and red lines indicate the upper and lower limits due to EBL fluctuations if the source is shifted to $z=0.3$, $0.5$, and $1.0$ and subsequently deabsorbed assuming the $\gamma$-rays travel through over and under dense regions along the entire path from the source to the observer.}
  \label{fig:mod}
\end{figure}
\begin{table}
  \caption{Deabsorbed slopes of blazar shifted to different $z$}
  \label{tab:mod}
  \begin{center}
    \begin{tabular}{ |c|c|c|c| }
      \hline
      Blazar      			& Redshift	& Observed $\Gamma$	&   Deabsorbed $\Gamma$        	      \\
      \hline
      \multirow{3}{*}{lES 0229+200} 	&  0.3		&  3.70			&   $1.72^{1.69}_{1.74}$ \vspace{0.1cm}\\
                                        &  0.5		&  5.56			&   $1.72^{1.66}_{1.77}$ \vspace{0.1cm}\\
                                        &  1.0		& 11.63			&   $1.72^{1.55}_{1.85}$ \vspace{0.1cm}\\
      \hline
    \end{tabular}
  \end{center}
\end{table}
\subsection{Impact on High Redshift $\gamma$-ray Spectra}
\label{ssc:hig}
As a final speculative outlook we calculate the maximum impact of the EBL fluctuations on $\gamma$-ray sources at larger redshifts than observed to date. As a starting point we use the deabsorbed spectrum of lES$0229+200$ (cf. Fig.~\ref{fig:slop}) which is a nearby high $\gamma$-ray energy source. We then shift this spectrum to various redshifts by ``absorbing'' it, i.e. multiplying it by $\exp[-\tau(z,E)_\text{mean}]$. This is the reverse process to deabsorbing it.

Figure~\ref{fig:CTA} shows the absorbed data points. Purple diamonds, red stars and green circles correspond to redshifts, $z =0.3$, $0.5$, and $1.0$, respectively. For comparison we copy the observed and the deabsorbed data as already displayed in Figure~\ref{fig:slop}. The grey line represents the sensitivity limit for $50h$ observations with Cherenkov Telescope Array (CTA). The third column of Table \ref{tab:mod} gives the  slopes of the absorbed (observable) data for the three redshifts. 

For $\gamma$-ray sources at increasing redshifts the high energy tails of the spectra drop below the detection limit at decreasing energies. The spectrum of the source shifted to $z=0.5$, for instance, drops below the CTA sensitivity for $\gamma$-ray energies above 700 GeV. It is interesting to note that the slope of the spectrum is very similar to the slope of the sensitivity limit. Thus a slight change in the slope of the observed spectrum, possibly due to fluctuations in the EBL, can have an effect on the detectability of the $\gamma$-ray source.

The 4th column of Table \ref{tab:mod} gives the slopes of the deabsorbed $\gamma$-ray spectra. Upper and lower limits are based on deabsorption for $\gamma$-rays propagating through over- or under-dense regions along the entire path from the source to the observer. Concretely, the absorbed (i.e. observable) data points are deabsorbed by multiplying with $\exp[\tau(z,E)_\text{mean}]$, $\exp[\tau(z,E)_\text{upper}]$ and $\exp[\tau(z,E)_\text{lower}]$. The deabsorbed data are fit be power laws. Figure~\ref{fig:mod} compares the ranges of the respective slopes to the intrinsic errors of the observation. The figure indicates that the change in the slope even for the most extreme cases is comparable to the intrinsic error of current measurements.
\section{Conclusion}
\label{sec:con}
Understanding the evolution of the EBL is crucial for the interpretation of $\gamma$-ray observations (and also for setting constraints on galaxy evolution models). We employ a simple model introduced by \cite{Razzaque2009} and include dust re-emission following \cite{Finke2010} to describe the evolution of the EBL spectrum. The use of oversimplified stellar evolution models results in spectra that are at the lower bound of current EBL measurements and theoretical predictions. This does not affect our conclusions since we are interested in \emph{relative differences} between EBL in high and low luminosity density environments. 

As an extension of current EBL models, we include EBL intensity fluctuations which are assumed to be the result of spatial fluctuations in the SFR. The statistical description of the spatial fluctuations of the SFR is derived from the semi-analytical galaxy catalogue based on MR7. The modified EBL model allows us to investigate the impact of EBL fluctuations on the attenuation of VHE $\gamma$-ray spectra due to  $\gamma + \gamma' \rightarrow e^+ + e^-$ pair production. As mentioned above, contributions of AGNs are ignored. However, despite the relatively small contribution of AGNs to the overall EBL spectrum the strong clustering of AGNs at high redshifts \citep[e.g.,][]{Porciani2004,Coil2007,Ross2009} may slightly enhance EBL fluctuations at early times. Testing this conjecture requires the implementation of an AGN model which is beyond the scope of the current work.

The overall outcome of our study is that, in general, the impact of EBL fluctuations on the slope of VHE $\gamma$-ray spectra is negligible. A deabsorption procedure which includes this effect deviates by less than $1\%$ from deabsorption based on mean EBL intensities.

\section*{Acknowledgments}
We thank the anonymous referee for providing constructive comments. This research was supported by the National Research Foundation (NRF). The Millennium Simulation (MR7) databases used in this paper and the web application providing online access to them were constructed as part of the activities of the German Astrophysical Virtual Observatory (GAVO).

\end{document}